\documentstyle[12pt]{article}
\newlength{\extraspace}
\newlength{\extraspaces}
\catcode`\@=11
\def\numberbysection{\@addtoreset{equation}{section}
\def\theequation{\arabic{section}.\arabic{equation}}}
\newcommand{\be}{\begin{equation}
\addtolength{\abovedisplayskip}{\extraspaces}
\addtolength{\belowdisplayskip}{\extraspaces}
\addtolength{\abovedisplayshortskip}{\extraspace}
\addtolength{\belowdisplayshortskip}{\extraspace}}
\newcommand{\ee}{\end{equation}}
\newcommand{\ba}{\begin{eqnarray}
\addtolength{\abovedisplayskip}{\extraspaces}
\addtolength{\belowdisplayskip}{\extraspaces}
\addtolength{\abovedisplayshortskip}{\extraspace}
\addtolength{\belowdisplayshortskip}{\extraspace}}
\newcommand{\ea}{\end{eqnarray}}
\newcommand{\newsection}[1]{
\vspace{7mm}
\pagebreak[3]
\addtocounter{section}{1}
\setcounter{equation}{0}
\setcounter{subsection}{0}
\setcounter{footnote}{0}
\begin{center}
{\large {\bf \thesection. #1}}
\end{center}
\nopagebreak
\medskip
\nopagebreak
\hspace{3mm}}
\newcommand{\nonu}{\nonumber \\[.5mm]}
\newcommand{\A}{&\!\!\!}

\newcommand{\R}{{\bf R}}
\newcommand{\tr}{\, {\rm tr}}
\newcommand{\e}{\, {\rm e}}

\newcommand{\FF}[2]{( F^{#1} F^{#2} )} 
\newcommand{\Ff}[2]{( F^{#1} \tilde F^{#2} )}

\renewcommand{\t}{\tilde}

\begin{document}
\addtolength{\baselineskip}{.7mm}
\thispagestyle{empty}
\begin{flushright}
STUPP--98--154 \\
{\tt hep-th/9808029} \\ 
August, 1998
\end{flushright}
\vspace{10mm}
\begin{center}
{\Large{\bf Duality Symmetries in Non-Linear \\[2mm]
Gauge Theories 
}} \\[25mm] 
{\sc Minako Araki}\footnote{
\tt e-mail: minako@th.phy.saitama-u.ac.jp} 
\hspace{2mm} and \hspace{2mm}
{\sc Yoshiaki Tanii}\footnote{
\tt e-mail: tanii@th.phy.saitama-u.ac.jp} \\[12mm]
{\it Physics Department, Faculty of Science \\[2mm]
Saitama University, Urawa, Saitama 338-8570, Japan} \\[25mm]
{\bf Abstract}\\[7mm]
{\parbox{13cm}{\hspace{5mm}
Duality symmetries are discussed for non-linear gauge theories 
of $(n-1)$-th rank antisymmetric tensor fields in general even 
dimensions $d=2n$. When there are $M$ field strengths and no 
scalar fields, the duality symmetry groups should be compact. 
We find conditions on the Lagrangian required by compact duality 
symmetries and show an example of duality invariant non-linear 
theories. We also discuss how to enlarge the duality symmetries 
to non-compact groups 
by coupling scalar fields described by non-linear sigma models. 
}}
\end{center}
\vfill
\newpage
\setcounter{section}{0}
\setcounter{equation}{0}
\numberbysection
%
%
\newsection{Introduction}
The free Maxwell's equations are invariant under SO(2) rotations 
of the electric field and the magnetic field into each other. 
The invariance of this kind is called the duality invariance. 
In the relativistic notation duality transformations in the 
Maxwell's theory are rotations of the electro-magnetic field 
strength $F_{\mu\nu}$ into its dual $\t F^{\mu\nu} = {1 \over 2} 
\epsilon^{\mu\nu\lambda\sigma} F_{\lambda\sigma}$. 
The duality invariance has generalizations in higher even 
dimensions. In $d=2n$ dimensions one can consider duality 
rotations of $n$-th rank field strengths of antisymmetric tensor 
fields and their duals. These duality symmetries naturally appear 
in supergravities \cite{FSZ}, \cite{SS}. Recently, duality 
symmetries play an important role in non-perturbative analysis 
of string theories \cite{HTW}. 
\par
In ref.\ \cite{GZ} the general discussion of duality symmetries 
in $d=4$ was given. It was shown that possible duality groups 
are Sp($2M$,\R) or its subgroup when there are $M$ field strengths, 
and that the Lagrangian is not invariant under duality 
transformations but should transforms in a definite way. 
Non-compact duality symmetries are possible when the theories 
contain scalar fields described by a G/H non-linear sigma model, 
where G is a non-compact group and H is a maximal compact subgroup. 
These results were generalized to theories in higher dimensions 
in ref.\ \cite{TANII}. It was found that possible duality 
groups are Sp($2M$,\R) or its subgroup in $d=4k$ and SO($M,\,M$) 
or its subgroup in $d=4k+2$. 
\par
More recently, duality symmetries were studied for non-linear 
gauge theories, whose Lagrangians are not quadratic in the 
field strengths of the gauge fields. Such Lagrangians appear 
as low energy effective theories in string theories. 
In refs.\  \cite{GR}, \cite{GZ2} conditions on the Lagrangians 
required by a compact duality symmetry U(1) were obtained 
when there is a single field strength in $d=4$. 
One of the theories satisfying the conditions is the 
Born-Infeld theory 
\be 
{\cal L}_{\rm BI} = {1 \over g^2} \left[ 1 - 
\sqrt{-{\rm det} \left( \eta_{\mu\nu} + g F_{\mu\nu} \right)} 
\right], 
\label{BI}
\ee
which appears as a low energy effective theory in open string 
theories. The compact duality symmetry U(1) can be enlarged to 
a non-compact symmetry Sp(2,\R) $\sim$ SL(2,\R) by coupling 
scalar fields (dilaton and axion fields). 
\par
The purpose of the present paper is to generalize these results 
on duality symmetries in non-linear gauge theories to theories 
which contain more than one field strengths, and also to theories 
in arbitrary even dimensions. (Theories with a single field strength 
in $d=4k$ dimensions were already studied in ref.\ \cite{GR}.) 
Such studies will be useful in discussing string dualities. 
When there are $M$ field strengths and no scalar fields, 
the duality symmetry group should be a compact group. 
We find conditions on the Lagrangian required by compact duality 
symmetries and show an example of duality invariant non-linear 
theories up to sixth order in the field strengths. 
We also discuss how to couple non-linear sigma models 
to enlarge the duality symmetry groups to non-compact ones. 
\par 
In the next section we review the general structure of duality 
symmetries in general even dimensions. In sect.\ 3 we work out 
the details of G/H non-linear sigma models for groups G, H 
relevant to duality symmetries. 
In sect.\ 4 conditions of duality invariance on the Lagrangian 
are obtained and an example of non-linear theories satisfying 
them is given. Finally, couplings to non-linear sigma models 
are discussed in sect. 5. 
%
%
\newsection{Duality symmetries}
In this section we review the general structure of duality 
symmetries in even dimensions $d=2n$ \cite{GZ}, \cite{TANII}. 
We consider theories of ($n-1$)-th rank antisymmetric tensor 
gauge fields $B^a_{\mu_1 \cdots \mu_{n-1}}(x)$ ($a = 1, \cdots, M$) 
interacting with other fields $\phi_i(x)$. 
Field strengths of the tensor gauge fields and their duals 
are defined as 
\ba
F^a_{\mu_1 \cdots \mu_n} 
\A = \A n \, \partial_{[\mu_1} B^a_{\mu_2 \cdots \mu_n]}, \nonu
\tilde F^{a\,\mu_1 \cdots \mu_n} 
\A = \A {1 \over n!} 
\epsilon^{\mu_1 \cdots \mu_n \nu_1 \cdots \nu_n} 
F^a_{\nu_1 \cdots \nu_n}, 
\ea
where the indices in $[\mu_1 \cdots \mu_n]$ are totally 
antisymmetrized with unit strength and 
$\epsilon^{\mu_1 \cdots \mu_d}$ is the Levi-Civita symbol. 
As the inclusion of gravity is straightforward, we consider 
theories in flat space-time. 
In $d$ dimensions the duality operation satisfies 
\be
\tilde{\! \tilde F} = \epsilon F, \qquad 
\epsilon = \cases{+1 & for \, $d=4k+2$, \cr
                  -1 & for \, $d=4k$. \cr } 
\ee
We consider a Lagrangian which is a function of $M$ 
field strengths and of other fields $\phi_i(x)$ 
and their derivatives 
\be
{\cal L} = {\cal L}(F^a_{\mu_1 \cdots \mu_n}, 
\phi_i, \partial_\mu \phi_i). 
\label{lagfund}
\ee
The gauge fields $B^a_{\mu_1 \cdots \mu_{n-1}}$ appear only 
through their field strengths $F^a_{\mu_1 \cdots \mu_n}$. 
\par
The equations of motion for $B^a_{\mu_1 \cdots \mu_{n-1}}$ 
and the Bianchi identities are 
\be
\partial_{\mu_1} \left( 
\tilde G^{a\,\mu_1 \cdots \mu_n} \right) = 0, \qquad
\partial_{\mu_1} \left( 
\tilde F^{a\,\mu_1 \cdots \mu_n} \right) = 0, 
\label{eqmotion}
\ee
where the duals of antisymmetric tensors 
$G^a_{\mu_1 \cdots \mu_n}$ are defined by 
\be
\tilde G^{a\,\mu_1 \cdots \mu_n} 
= {n! 
} \, {\partial {\cal L} \over \partial 
F^a_{\mu_1 \cdots \mu_n}}. 
\label{gdef}
\ee
For the free Maxwell theory in four dimensions, 
we obtain $G^{\mu\nu} = \tilde F^{\mu\nu}$. 
Eq.\ (\ref{eqmotion}) is invariant under transformations 
\be
\delta \left(
\begin{array}{c}
G \\
F 
\end{array}
\right) 
= 
\left(
\begin{array}{cc}
A & B \\
C & D 
\end{array}
\right)
\left(
\begin{array}{c}
G \\
F 
\end{array}
\right), \qquad \delta \phi^i = \xi^i(\phi), 
\label{gltrans}
\ee
where $A$, $B$, $C$, $D$ are constant $M \times M$ real matrices 
and $\xi^i(\phi)$ are functions of $\phi^i$. 
These constants are not independent and should satisfy certain 
conditions obtained from the covariance of the definition of $G$ 
(\ref{gdef}) and the covariance of the equations of motion 
for $\phi^i$ under the transformations (\ref{gltrans}). 
\par
Let us first consider the covariance of the definition of $G$. 
By eq.\ (\ref{gdef}) $G$ is given as functions of 
$F$ and $\phi$. Therefore, the transformation of $G$ can be 
derived from those of $F$ and $\phi$. From eqs.\ (\ref{gdef}) 
and (\ref{gltrans}) we obtain 
\be
\delta \tilde G^a 
= n! \,{\partial \delta {\cal L} \over \partial F^a} 
- \tilde G^b D^{ba} 
- \tilde G^b C^{bc} {\partial G^c \over \partial F^a}. 
\ee
This should be consistent with the transformation of $G$ 
given in eq.\ (\ref{gltrans}). 
By equating these two transformation laws we obtain 
\ba
\A \A {\partial \over \partial F^a} \left( 
n! \delta {\cal L} - {1 \over 2} F^b B^{bc} \tilde F^c 
- {1 \over 2} \tilde G^b C^{bc} G^c \right) 
- \left(  A^{ab} + D^{ba} \right) 
n! {\partial {\cal L} \over \partial F^b} \nonu
\A \A \qquad\qquad\qquad
= {1 \over 2} \left( B^{ab} + \epsilon B^{ba} \right) \tilde F^b 
+ {1 \over 2} \tilde G^b \left( C^{bc} + \epsilon C^{cb} \right) 
{\partial G^c \over \partial F^a}. 
\ea
When there exist non-trivial interactions, this equation gives 
conditions on the transformation parameters 
\be
A^{ab} + D^{ba} = \alpha \delta^{ab}, \qquad
B^{ab} = - \epsilon B^{ba}, \qquad
C^{ab} = - \epsilon C^{ba}, 
\label{conditionone}
\ee
where $\alpha$ is an arbitrary constant, as well as a condition 
on the variation of the Lagrangian 
\be
{\partial \over \partial F^a} \left(
\delta {\cal L} - {1 \over 2n!} F^a B^{ab} \tilde F^b 
- {1 \over 2n!} \tilde G^a C^{ab} G^b 
- \alpha {\cal L} \right) = 0. 
\label{conditiontwo}
\ee
The equations of motion for $\phi^i$ are 
\be
E_i \equiv 
\left(
{\partial \over \partial \phi^i} - \partial_\mu 
{\partial \over \partial (\partial_\mu \phi^i)} \right)
{\cal L} = 0. 
\ee
The covariance of these equations under the duality 
transformation (\ref{gltrans}) 
\be
\delta E_i = - {\partial \xi^j \over \partial \phi^i} E_j 
\ee
requires another condition on the variation of the Lagrangian 
\be
\left(
{\partial \over \partial \phi^i} - \partial_\mu 
{\partial \over \partial (\partial_\mu \phi^i)} \right) 
\left( \delta {\cal L} - {1 \over 2n!} 
\tilde G^a C^{ab} G^b \right) = 0. 
\label{conditionthree}
\ee
\par
We can now find out possible duality groups from 
eqs.\ (\ref{conditionone}), (\ref{conditiontwo}) and 
(\ref{conditionthree}). Comparing eqs.\ (\ref{conditiontwo}) 
and (\ref{conditionthree}) we find $\alpha = 0$. 
Then, the conditions on the parameters (\ref{conditionone}) can 
be written as 
\be
X^T \Omega + \Omega X = 0, 
\label{xomega}
\ee
where 
\be
X = 
\left(
\begin{array}{cc}
A & B \\
C & D 
\end{array}
\right), 
\qquad
\Omega = 
\left(
\begin{array}{cc}
0 & \epsilon \\
1 & 0 
\end{array}
\right). 
\label{ncgroup}
\ee
For $d = 4k$ ($\epsilon = -1$) $\Omega$ is an antisymmetric matrix 
and the above condition implies that the group of duality 
transformations is Sp($2M$,\,\R) or its subgroup. 
On the other hand, for $d = 4k+2$ ($\epsilon = +1$) $\Omega$ is a 
symmetric matrix, which can be diagonalized 
to diag$({\bf 1}, -{\bf 1})$. The group of duality 
transformations in this case is SO($M$,\,$M$) or its subgroup. 
Eqs.\ (\ref{conditionone}), (\ref{conditiontwo}) and 
(\ref{conditionthree}) also require that the Lagrangian 
must transform as 
\ba
\delta {\cal L} 
\A = \A {1 \over 2n!} F^a B^{ab} \tilde F^b 
+ {1 \over 2n!} \tilde G^a C^{ab} G^b \nonu
\A = \A \delta \left( {1 \over 2n!} F^a \tilde G^a \right). 
\label{ltransf}
\ea
Thus, although the Lagrangian is not invariant under the duality 
transformations, it transforms in a definite way. 
It can be shown that a derivative of the Lagrangian with 
respect to an invariant parameter is invariant 
under the duality transformations. 
The invariant parameter can be an invariant external field 
such as the metric. 
Thus, the energy-momentum tensor, which is given by a functional 
derivative of the Lagrangian with respect to the metric, 
is invariant under the duality transformations. 
\par
In field theories with non-compact symmetries one has to 
check the absence of ghosts, i.e., negative norm states. 
When no scalar fields are present, only compact subgroups of 
Sp($2M$,\,\R) or SO($M$,\,$M$) are possible as duality 
symmetries without introducing ghosts. The maximal compact 
subgroup of Sp($2M$,\,\R) and SO($M$,\,$M$) are U($M$) 
and SO($M$) $\times$ SO($M$) respectively. 
Elements of these compact groups are given by $X$ in 
eq.\ (\ref{ncgroup}) with 
\be
D = A, \quad C = \epsilon B, 
\quad A^T = -A, \quad B^T = -\epsilon B. 
\label{compact}
\ee
Non-compact duality symmetries are possible when scalar fields 
described by a G/H non-linear sigma model are present. 
Here, G is a non-compact duality group and H is a maximal 
compact subgroup of G. In the next section we work out the 
details of G/H non-linear sigma models for G = Sp($2M$,\,\R), 
H = U($M$) in $d=4k$ and G = SO($M$,\,$M$), 
H = SO($M$) $\times$ SO($M$) in $d=4k+2$. 
%
%
\newsection{Non-linear sigma models}
The G/H non-linear sigma model \cite{CWZ}, \cite{GZ} is a theory 
of G/H-valued scalar fields, which is invariant under rigid G 
transformations. For our purpose G is a non-compact Lie group 
and H is a maximal compact subgroup of G. 
We represent the scalar fields by a G-valued scalar field $V(x)$ 
and require local H invariance. Since we do not introduce independent 
H gauge fields, the H part of $V(x)$ can be gauged away and physical 
degrees of freedom are on a coset space G/H. 
Under the rigid G transformation $g$ and the local H 
transformation $h(x)$ the scalar field $V(x)$ transforms as 
\be
V(x) \rightarrow g V(x) h^{-1}(x). 
\label{ghtrans}
\ee
\par
To construct the action we decompose the Lie algebra ${\bf G}$ 
of G as ${\bf G} = {\bf H} + {\bf N}$, where ${\bf H}$ is the 
Lie algebra of H and ${\bf N}$ is its orthogonal complement 
in ${\bf G}$. The orthogonality is defined with respective to 
the trace in a certain representation: 
$\tr ({\bf H} \, {\bf N}) = 0$. 
The ${\bf G}$-valued field $V^{-1} \partial_\mu V$ is 
decomposed as 
\be
V^{-1} \partial_\mu V = Q_\mu + P_\mu, \qquad 
Q_\mu \in {\bf H}, \quad P_\mu \in {\bf N}. 
\label{pqdecom}
\ee
The transformation laws of $Q_\mu$ and $P_\mu$ under the local 
H transformation in (\ref{ghtrans}) are found to be 
\be
Q_\mu \rightarrow h Q_\mu h^{-1} + h \partial_\mu h^{-1}, \qquad
P_\mu \rightarrow h P_\mu h^{-1}, 
\ee
while they are invariant under the rigid G 
transformations. We see that $Q_\mu$ transforms 
as an H gauge field, while $P_\mu$ is covariant under 
the H transformations. 
By using these quantities we can construct actions which are 
invariant under the rigid G and the local H transformations. 
The kinetic term of the scalar fields is 
\be
{\cal L}_{\rm scalar} = {1 \over 2} \tr ( P_\mu P^\mu ), 
\label{nsmlag}
\ee
which is quadratic in derivatives of $V$ and is manifestly 
invariant under the rigid G and the local H transformations. 
The H gauge field $Q_\mu$ can be used to define the covariant 
derivatives on other fields transforming under the local H. 
The covariant quantity $P_\mu$ also can be used to construct 
invariant terms in the action. 
\par
Let us work out the details of the theories relevant to duality 
symmetries. For G = Sp($2M$,\,\R), H = U($M$) in $d=4k$ 
the G transformations in the complex basis are 
\be
\left(
\begin{array}{c}
F + i G \\
F - i G 
\end{array}
\right) 
\rightarrow 
\left(
\begin{array}{cc}
a & b^* \\
b & a^* 
\end{array}
\right) 
\left(
\begin{array}{c}
F + i G \\
F - i G 
\end{array}
\right), 
\quad
a^\dagger a - b^\dagger b = 1, \quad 
a^T b - b^T a = 0, 
\ee
where $a$, $b$ are $M \times M$ complex matrices. 
The subgroup H = U($M$) corresponds to the case $b = 0$. 
In this basis the G-valued scalar field is expressed as 
\be
V(x) = 
\left(
\begin{array}{cc}
\phi_0(x) & \phi_1^*(x) \\
\phi_1(x) & \phi_0^*(x)
\end{array}
\right), 
\quad
\phi_0^\dagger \phi_0 - \phi_1^\dagger \phi_1 = 1, \quad 
\phi_0^T \phi_1 - \phi_1^T \phi_0 = 0. 
\label{4kscalar}
\ee
The physical degrees of freedom are represented by 
an H-invariant variable 
\be
z = \phi_1^* (\phi_0^*)^{-1} = z^T, 
\ee
which transforms under G as 
\be
z \rightarrow (a z + b^*) (b z + a^*)^{-1}. 
\label{4kztrans}
\ee
Alternatively, we can use a variable 
\be
S = S_1 + i S_2 = i {1-z \over 1+z}, 
\qquad S_1, S_2 \in {\bf R} 
\label{4ks}
\ee
to represent physical degrees of freedom. 
Its infinitesimal G transformation is 
\be
\delta S = A S + B - S C S - S D. 
\label{4kstrans}
\ee
The kinetic term for 
the scalar fields (\ref{nsmlag}) becomes 
\ba
{\cal L}_{\rm scalar} 
\A = \A \tr \left[ \left( \phi_0^T \partial_\mu \phi_1 
- \phi_1^T \partial_\mu \phi_0 \right)^* 
\left( \phi_0^T \partial^\mu \phi_1 
- \phi_1^T \partial^\mu \phi_0 \right) \right] \nonu
\A = \A \tr \left[ {1 \over 1 - z^* z} \partial_\mu z^* 
{1 \over 1 - z z^*} \partial^\mu z \right] \nonu
\A = \A - \tr \left[ {1 \over S-S^*} 
\partial_\mu S^* {1 \over S-S^*} 
\partial^\mu S \right]. 
\label{4kaction}
\ea
\par
We now turn to the case G = SO($M$,\,$M$), 
H = SO($M$) $\times$ SO($M$) in $d=4k+2$. In the 
$\Omega$-diagonal basis the G transformations are written as 
\ba
\A\A \left(
\begin{array}{c}
F + G \\
F - G 
\end{array}
\right) 
\rightarrow 
\left(
\begin{array}{cc}
a & b \\
c & d 
\end{array}
\right) 
\left(
\begin{array}{c}
F + G \\
F - G 
\end{array}
\right), \nonu
\A\A \qquad
a^T a - c^T c = 1, \quad 
d^T d - b^T b = 1, \quad 
a^T b - c^T d = 0, 
\ea
where $a$, $b$, $c$, $d$ are $M \times M$ real matrices. 
The subgroup H = SO($M$) $\times$ SO($M$) corresponds to the 
case $b = c = 0$. The G-valued scalar field is parametrized as 
\ba
\A\A V(x) = 
\left(
\begin{array}{cc}
\phi_1(x) & \psi_2(x) \\
\psi_1(x) & \phi_2(x)
\end{array}
\right), \nonu
\A\A \qquad
\phi_1^T \phi_1 - \psi_1^T \psi_1 = 1, \quad 
\phi_2^T \phi_2 - \psi_2^T \psi_2 = 1, \quad 
\phi_1^T \psi_2 - \psi_1^T \phi_2 = 0. 
\ea
We define an H-invariant variable 
\be
z = ( \psi_1 (\phi_1)^{-1} )^T = \psi_2 (\phi_2)^{-1}, 
\ee
which transforms under the G transformation as 
\be
z \rightarrow (az+b)(cz+d)^{-1}. 
\label{4k2ztrans}
\ee
Alternatively, we can use a variable 
\be
S = - S_1 + S_2 = {1-z \over 1+z}, \qquad 
S_1^T = - S_1, \quad S_2^T = S_2, 
\label{4k2s}
\ee
whose infinitesimal G transformation is 
\be
\delta S = A S - B + S C S - S D.
\label{4k2strans}
\ee
The kinetic term (\ref{nsmlag}) then becomes 
\ba
{\cal L}_{\rm scalar} 
\A = \A \tr \left[ \left( \phi_1^T \partial_\mu \psi_2 
- \psi_1^T \partial_\mu \phi_2 \right) 
\left( \phi_2^T \partial^\mu \psi_1 
- \psi_2^T \partial^\mu \phi_1 \right) \right] \nonu
\A = \A \tr \left[ {1 \over 1 - z^T z} \partial_\mu z^T 
{1 \over 1 - z z^T} \partial^\mu z \right] \nonu
\A = \A \tr \left[ {1 \over S+S^T} \partial_\mu S^T 
{1 \over S+S^T} \partial^\mu S \right]. 
\label{4k2action}
\ea
%
%
\newsection{Non-linear gauge theories}
Let us consider duality invariant gauge theories, 
which are not quadratic in the field strengths. 
In this section we shall first consider theories which 
contain only gauge fields and no other fields. 
The duality symmetry group should be a compact group. 
We consider the maximal symmetric cases: U($M$) for $d=4k$ 
and SO($M$) $\times$ SO($M$) for $d=4k+2$. 
The transformation parameters satisfy eq.\ (\ref{compact}). 
\par
According to the general discussion in sect.\ 2 the Lagrangian 
must satisfies eq.\ (\ref{ltransf}). 
On the other hand, assuming that the Lagrangian is a function 
of only the field strengths $F^a$ we obtain 
\be
\delta {\cal L} 
= \delta F^a {\partial {\cal L} \over \partial F^a} 
= {1 \over n!} \left( C^{ab} G^b + D^{ab} F^b \right) \tilde G^a. 
\ee
By equating these two expressions and by using 
eq.\ (\ref{compact}) we obtain the conditions 
\ba 
F^{a \mu_1 \cdots \mu_n} \t F^b_{\mu_1 \cdots \mu_n} 
+ G^{a \mu_1 \cdots \mu_n} \t G^b_{\mu_1 \cdots \mu_n}
\A=\A 0, 
\label{FGcond} \\
F^{a \mu_1 \cdots \mu_n} \t G^b_{\mu_1 \cdots \mu_n} 
- F^{b \mu_1 \cdots \mu_n} \t G^a_{\mu_1 \cdots \mu_n}
\A=\A 0. 
\label{FGcond2} 
\ea 
The first condition (\ref{FGcond}) is a generalization of 
the condition for a single gauge field obtained in 
refs.\ \cite{GR}, \cite{GZ2}. From eq.\ (\ref{gdef}) the second 
condition (\ref{FGcond2}) can be rewritten as 
\be
F^a {\partial {\cal L} \over \partial F^b} 
- F^b {\partial {\cal L} \over \partial F^a} = 0. 
\ee
We see that it requires the SO($M$) invariance of the Lagrangian 
when $F^a$ transform as an $M$-dimensional vector of SO($M$). 
\par
Let us obtain an example of the Lagrangians which satisfy these 
conditions. We expand the Lagrangian in $F$ and consider terms 
up to and including of order ${\cal O}(F^6)$ 
\be
{\cal L} = {\cal L}_2 + {\cal L}_4 + {\cal L}_6 + {\cal O}(F^8), 
\label{Lagexp}
\ee
where ${\cal L}_m$ is of order ${\cal O}(F^m)$. 
We make the following ansatz 
\ba
{\cal L}_2 \A = \A \alpha_1 ( F^a F^a ) 
+ \alpha_2 \Ff{a}{a}, \nonu
{\cal L}_4 \A = \A \beta_1 (F^a F^a)^2 
+ \beta_2 (F^a F^b)^2 
+ \beta_3 (F^a \tilde F^a )^2 
+ \beta_4 (F^a \tilde F^b )^2, \nonu
{\cal L}_6 \A = \A
\gamma_1 ( F^a F^a )^3 
+ \gamma_2 ( F^a F^b )^2 ( F^c F^c ) 
+ \gamma_3 ( F^a F^b ) ( F^a F^c ) ( F^b F^c ) \nonu 
\A \A 
+ \gamma_4 ( F^a \tilde F^a )^2 ( F^c F^c ) 
+ \gamma_5 ( F^a \tilde F^b )^2 ( F^c F^c ) \nonu 
\A \A
+ \gamma_6 ( F^a \tilde F^b ) ( F^a F^b ) ( F^c \tilde F^c ) 
+ \gamma_7 ( F^a \tilde F^b ) ( F^a \tilde F^c ) ( F^b F^c ), 
\label{L246}
\ea
where $( F^a F^b ) 
= F^a_{\mu_1 \cdots \mu_n} F^{b \mu_1 \cdots \mu_n}$ etc. 
We note that the terms proportional to 
$\Ff{a}{a}$ identically vanish in $d = 4k+2$ 
since $\Ff{a}{a} = - (\t F^a F^a) = 0$. 
Since the indices $a, b, \cdots$ are appropriately contracted, 
the condition (\ref{FGcond2}) is automatically satisfied. 
Substituting eqs.\ (\ref{Lagexp}), (\ref{L246}) into 
eq.\ (\ref{FGcond}) and using the identities 
\be
(F_1 \tilde F_2) = - \epsilon (\tilde F_1 F_2), \qquad 
(\tilde F_1 \tilde F_2) = - (F_1 F_2), 
\ee
we find that the condition (\ref{FGcond}) is 
satisfied at this order when the non-vanishing parameters are 
\be
\alpha_1 = \pm \frac{1}{2n!}, \qquad 
\beta_2 = \beta_4 = \alpha_1 \beta, \qquad
\gamma_3 = \gamma_7 = 2 \alpha_1 \beta^2. 
\ee
Thus the Lagrangian up to sixth order in $F$ is 
\ba 
{\cal L} \A = \A \pm {1 \over 2n!} \biggl[ \FF{a}{a} 
+ \beta \left\{ \FF{a}{b} \FF{a}{b} 
+ \Ff{a}{b} \Ff{a}{b} \right\} \nonu 
\A \A 
+ 2 \beta^2 \left\{ 
\FF{a}{b} \FF{a}{c} \FF{b}{c} 
+ \Ff{a}{b} \Ff{a}{c} \FF{b}{c} \right\} 
\biggr] + {\cal O}(F^8). 
\label{upto6th}
\ea
For $M$=1 in four dimensions we see that this Lagrangian 
coincides with the Born-Infeld Lagrangian (\ref{BI}) up to 
this order. 
%
%
\newsection{Coupling to scalar fields}
The compact duality symmetries in the previous section can be 
enlarged to non-compact symmetries by introducing scalar 
fields. In this section we shall discuss couplings of scalar 
fields described by G/H non-linear sigma models to non-linear 
gauge theories. We require that the theories are invariant 
under the maximal non-compact duality symmetries 
Sp($2M$,\,\R) in $d=4k$ or SO($M$,\,$M$) in $4k+2$. 
\par
We use the fields $S_1$, $S_2$ introduced in eqs.\ (\ref{4ks}), 
(\ref{4k2s}) to represent the G/H-valued scalar fields. 
They have symmetry properties $S_1^T = - \epsilon S_1$, 
$S_2^T = S_2$. It is convenient to introduce 
the ``vielbein'' $R^{ab}$ for the ``metric'' $S^{ab}_2$ by 
\be
S_2 = R^T R. 
\ee
For a given $S_2$, $R$ is determined only up to local SO($M$) 
transformations 
\be
\delta R(x) = \Lambda(x) R(x), \qquad
\Lambda^T(x) = - \Lambda(x). 
\label{localso}
\ee
When the Lagrangian is expressed by using $R$, it should be 
invariant under these local SO($M$) transformations. 
The duality transformations of $S_1$, $S_2$ can be derived from 
eqs.\ (\ref{4kstrans}), (\ref{4k2strans}) 
\ba 
\delta S_1 \A = \A 
A S_1 + B - S_1 C S_1 - \epsilon R^T R C R^T R - S_1 D, \nonu 
\delta R \A = \A - R C S_1 - R D, 
\label{srtrans}
\ea 
where the transformation parameters satisfy eq.\ (\ref{xomega}). 
\par 
Let us obtain conditions for the duality invariance of the 
theory. Since the scalar kinetic term (\ref{4kaction}) or 
(\ref{4k2action}) is invariant under the duality transformations 
(\ref{srtrans}) by itself, we consider the Lagrangian 
${\cal L}$ excluding it. Then, we can assume that ${\cal L}$ 
does not depend on derivatives of the scalar fields. 
The Lagrangian must satisfies the transformation property 
(\ref{ltransf}). On the other hand, the variation can be 
expressed also as 
\be
\delta {\cal L} 
= \delta F^a {\partial {\cal L} \over \partial F^a} 
+ \delta S_1^{ab} {\partial {\cal L} \over \partial S_1^{ab}} 
+ \delta R^{ab} {\partial {\cal L} \over \partial R^{ab}}, 
\ee
where $\delta F^a$ and $\delta S_1$, $\delta R$ are given in 
eqs.\ (\ref{gltrans}) and (\ref{srtrans}) respectively. 
By equating these two expressions we obtain the conditions 
on the Lagrangian. From the coefficients of $B^{ab}$ we obtain 
\be
{\partial {\cal L} \over \partial S_1^{ab}} 
= {1 \over 2n!} F^a \tilde F^b. 
\ee
This condition can be solved by 
\be
{\cal L} = {1 \over 2n!} F^a S_1^{ab} \tilde F^b 
+ \bar{\cal L}(F, R), 
\ee
where $\bar{\cal L}(F, R)$ is independent of $S_1$. 
Next, from the coefficients of $A^{ab} = - D^{ba}$ we obtain 
\be
F^a {\partial \bar{\cal L} \over \partial F^b} 
- R^{cb} {\partial \bar{\cal L} \over \partial R^{ca}} = 0. 
\label{adcondition}
\ee
To solve this condition we introduce 
\be
{\cal F}^a = R^{ab} F^b, 
\ee
and regard $\bar{\cal L}$ as a function of ${\cal F}$ and $R$. 
By using identities 
\ba
{\partial \bar{\cal L}(F, R) \over \partial F^b} 
\A = \A R^{cb} {\partial \bar{\cal L}({\cal F}, R) 
\over \partial {\cal F}^c}, \nonu
{\partial \bar{\cal L}(F, R) \over \partial R^{ca}} 
\A = \A 
{\partial \bar{\cal L}({\cal F}, R) \over \partial R^{ca}} 
+ F^a {\partial \bar{\cal L}({\cal F}, R) 
\over \partial {\cal F}^c}, 
\ea
eq.\ (\ref{adcondition}) becomes 
\be
{\partial \bar{\cal L}({\cal F}, R) \over \partial R^{ab}} = 0. 
\ee
Thus, $\bar{\cal L}({\cal F}, R) = \bar{\cal L}({\cal F})$ 
is independent of $R$. 
Finally, from the coefficients of $C^{ab}$ we obtain a condition 
\be
{\cal F}^a \tilde{\cal F}^b + {\cal G}^a \tilde{\cal G}^b = 0, 
\label{ncFGcond}
\ee
where we have defined 
\be
\tilde{\cal G}^a 
= n! {\partial \bar{\cal L}({\cal F}) 
\over \partial {\cal F}^a}. 
\ee
We must also require the invariance of the Lagrangian under the 
local SO($M$) transformations (\ref{localso}). 
It gives a condition 
\be
{\cal F}^a \tilde{\cal G}^b - {\cal F}^b \tilde{\cal G}^a = 0. 
\label{ncFGcond2}
\ee
\par
To summarize the duality invariance requires that the Lagrangian 
takes the form 
\be
{\cal L} = {\cal L}_{\rm scalar}(S, \partial_\mu S) 
+ {1 \over 2n!} F^a S_1^{ab} \tilde F^b 
+ \bar{\cal L}({\cal F}), 
\label{ncaction}
\ee
where the scalar kinetic term ${\cal L}_{\rm scalar}$ given 
in eq.\ (\ref{4kaction}) or (\ref{4k2action}) is included 
and $\bar{\cal L}({\cal F})$ should satisfy 
eqs.\ (\ref{ncFGcond}), (\ref{ncFGcond2}). 
The conditions (\ref{ncFGcond}), (\ref{ncFGcond2}) 
take the same form as eqs.\ (\ref{FGcond}), (\ref{FGcond2}). 
Therefore, once we find a theory containing only gauge fields 
which has a compact duality symmetry, we can construct a theory 
which has non-compact duality symmetry by coupling it to scalar 
fields as in eq.\ (\ref{ncaction}). 
We note that the transformations of ${\cal F}$ and ${\cal G}$ 
have a simple form 
\ba
\delta {\cal F}^a \A = \A 
(R C R^T)^{ab} {\cal G}^b + \Lambda^{ab} {\cal F}^b, \nonu
\delta {\cal G}^a \A = \A 
\epsilon (R C R^T)^{ab} {\cal F}^b + \Lambda^{ab} {\cal G}^b. 
\label{caltrn} 
\ea
These have the same form as the compact U($M$) or 
SO($M$) $\times$ SO($M$) transformations in eq.\ (\ref{compact}) 
but with field dependent and local transformation parameters. 
\par 
For $M=1$ in four dimensions these results reduce to those in 
refs.\ \cite{GR}, \cite{GZ2}. In particular, the scalar fields 
have only two components $S_1 = a$ and 
$R = \e^{-{1 \over 2}\phi}$, where $a(x)$ 
and $\phi(x)$ are the axion field and the dilaton field 
respectively. The Lagrangian (\ref{ncaction}) then becomes 
\be
{\cal L} = {1 \over 2} (\partial_\mu \phi)^2 
+ {1 \over 2} \e^{2 \phi} (\partial_\mu a)^2 
+ {1 \over 4} a F_{\mu\nu} \tilde F^{\mu\nu} 
+ \bar{\cal L}(\e^{-{1 \over 2}\phi} F_{\mu\nu}), 
\ee
where $\bar{\cal L}$ must satisfies eq.\ (\ref{ncFGcond}). 
The condition (\ref{ncFGcond2}) is automatically satisfied 
for $M=1$. 

\vspace{10mm}

\noindent Note added

After completing this paper we noticed ref.\ \cite{CJLP}, 
in which duality symmetries of supergravities in general 
space-time dimensions were discussed in considerable details. 
We would like to thank B. Julia for informing us about 
these works. 

\newpage
%
%
\newcommand{\NP}[1]{{\it Nucl.\ Phys.\ }{\bf #1}}
\newcommand{\PL}[1]{{\it Phys.\ Lett.\ }{\bf #1}}
\newcommand{\CMP}[1]{{\it Commun.\ Math.\ Phys.\ }{\bf #1}}
\newcommand{\MPL}[1]{{\it Mod.\ Phys.\ Lett.\ }{\bf #1}}
\newcommand{\IJMP}[1]{{\it Int.\ J. Mod.\ Phys.\ }{\bf #1}}
\newcommand{\PR}[1]{{\it Phys.\ Rev.\ }{\bf #1}}
\newcommand{\PRL}[1]{{\it Phys.\ Rev.\ Lett.\ }{\bf #1}}
\newcommand{\PTP}[1]{{\it Prog.\ Theor.\ Phys.\ }{\bf #1}}
\newcommand{\PTPS}[1]{{\it Prog.\ Theor.\ Phys.\ Suppl.\ }{\bf #1}}
\newcommand{\AP}[1]{{\it Ann.\ Phys.\ }{\bf #1}}
\end{document}